\documentclass[showpacs,aps,amssymb,pra,amsmath]{revtex4}
\usepackage{epsfig}
\usepackage{bm}
\usepackage{amsmath,graphicx}
\usepackage{subfig}
\usepackage{setspace}
\usepackage{color}

\newcommand{\beq}{\begin{equation}}
\newcommand{\eeq}{\end{equation}}
\newcommand{\bea}{\begin{eqnarray}}
\newcommand{\eea}{\end{eqnarray}}
\def\bk{{\bf k}}
\def\bq{{\bf q}}

\def\br{{\bf r}}
\def\bp{{\bf p}}
\def\bs{{\bf s}}
\def\bR{{\bf R}}

\begin{document}
\title{Thomas-Fermi von Weizs\"acker theory for a harmonically
trapped, two-dimensional, spin-polarized dipolar Fermi gas}
\author{B. P. van Zyl}
\affiliation{Department of Physics, St. Francis Xavier University, Antigonish, NS,
B2G 2W5}
\author{E. Zaremba and P. Pisarski}
\affiliation{Department of Physics, Astronomy and Engineering
Physics, Queen's University, Kingston, Ontario 
 K7L 3N6, Canada.}
\date{\today}
\begin{abstract}
We systematically develop a 
density functional description for the equilibrium
properties of a two-dimensional, harmonically trapped,
spin-polarized dipolar Fermi gas based on the Thomas-Fermi von Weizs\"acker approximation.
We pay particular attention
to the construction of the two-dimensional kinetic energy
functional, where corrections beyond the local
density approximation must be motivated with care. We also
present an intuitive derivation of the interaction energy functional 
associated with the dipolar interactions, and
provide physical insight into why it can be represented as a
local functional. Finally, a simple, and highly efficient
self-consistent numerical procedure is developed to determine
the equilibrium density of the system for a range of dipole 
interaction strengths.

\end{abstract}
\maketitle

\section{Introduction}
Ultra-cold, trapped dipolar quantum gases have received
increasing
attention over the past decade owing to the inherently interesting 
properties of the anisotropic, and long-range
nature of the dipole-dipole interaction~\cite{baranov}. One of 
the important consequences of the anisotropy is
that the interactions between the particles can be tuned from
being predominantly attractive to repulsive by simply
changing the 3D trapping geometry, or for dipoles confined to 
the 2D $x$-$y$ plane, by adjusting
the orientation of the dipoles relative to the
$z$-axis~\cite{baranov,bruun}.
Therefore, novel physics in
both the equilibrium and dynamic properties of such systems may be 
explored as a function of the strength of the interaction, the geometry of the confining potential, and
the dimensionality of the system.

While the degenerate dipolar Bose gas has been well studied 
experimentally and theoretically~\cite{baranov}, realizing a degenerate dipolar 
Fermi gas in the laboratory has 
proven to be much more elusive. One of the reasons
for this is that the path to quantum degeneracy is impeded by
the Pauli principle, which forbids $s$-wave collisions
between identical atoms. Thus, early attempts to cool both
magnetic and molecular dipolar Fermi gases below degeneracy were
unsuccessful~\cite{chic,berg,ni,chotia}.  However, in the recent 
work of M.~Lu {\it et al.}~\cite{lu}, this experimental hurdle was 
finally overcome, resulting in the experimental realization of 
a spin polarized, degenerate dipolar Fermi gas.
Specifically, using the method of sympathetic cooling, 
a mixture consisting of $^{161}$Dy and 
the bosonic isotope $^{162}$Dy, were cooled to $T/T_F\sim 0.2$. 
In addition, this group was also able to evaporatively cool a single
component gas of $^{161}$Dy down to a temperature of $T/T_F \sim 0.7$.
This latter result is presumed to arise from the rethermalization
provided by the strong dipolar
scattering between the $^{161}$Dy atoms which have a large
magnetic moment ($\mu \sim 10 \mu_B$).

The ability to fabricate such systems in the laboratory now opens the door for
the investigation of both the equilibrium and dynamical properties
of dipolar Fermi gases, and will enable contact to be made with the
large body of theoretical work already in the
literature~\cite{baranov}. Moreover, it is now reasonable to 
expect that quasi-2D degenerate dipolar Fermi gases will also be 
realized experimentally,
thereby allowing for studies into 
the stability, pairing, and superfluidity of low-dimensional 
dipolar systems, which to date have only been investigated
theoretically~\cite{bruun,seiberer,baranov2}.

With a view towards ultimately calculating the collective 
mode frequencies, we develop in this paper
a density-functional theory (DFT) for the equilibrium properties of a 
degenerate, harmonically trapped, spin polarized dipolar Fermi gas. Our theoretical framework is based on the
Thomas-Fermi von Weizs\"acker (TFvW) approximation which was
previously formulated in the context of degenerate electron
gases~\cite{zaremba_tso}.
The mathematical framework of the TFvW theory is very simple,
numerically easy to implement, and computationally inexpensive. 
The TFvW theory has also been shown to provide an exceedingly 
accurate description of equilibrium properties, as well as
collective excitations ({\it i.e.,} magnetoplasmons), of electronic 
systems in a variety of two-and three-dimensional confinement
geometries~\cite{PhD,vanzyl1,vanzyl2,vanzyl3,vanzyl4,zaremba_tso,zaremba96}.
Our purpose here is to take advantage of this approach, which 
is largely unknown in the cold-atoms community, and apply it to 
the dipolar Fermi gas. 
We will only address the equilibrium properties in
this paper, and leave the presentation of the more involved mode 
calculation to a future publication. Moreover, in anticipation of 
forthcoming experiments, along with the goal of making contact 
with the recent theoretical work of Fang and
Englert~\cite{fang}, we will focus on the 2D 
geometry, although the extension of the theory to 3D is 
straightforward. The 2D geometry also allows for the development
of exact analytical results, which we will exploit to test the
efficacy of the TFvW approximation.

The organization of our paper is as follows.  In Sec. II, 
we construct the approximate kinetic energy functional for the trapped 
2D system, 
and show that it is neccessary to go beyond the local-density 
approximation (LDA) in order to provide a more accurate determination 
of the ground state energy, along with 
physically reasonable density profiles, for the system.  In Sec. III, 
we present an intuitive approach for the
development of the interaction energy functional, in addition to providing
physical insight into why the functional may be represented soley in 
terms of the {\em local} density.
Section IV presents the details of the numerical procedure for 
implementing the TFvW, in addition to representative illustrations of 
the spatial density profile of the dipolar
gas as the dipole-dipole interaction strength is changed.  Finally, in Sec. V, 
we present our conclusions and closing remarks.

\section{Kinetic Energy Functional}
In the Kohn-Sham (KS) 
DFT~\cite{KS}, the 
noninteracting kinetic energy is treated exactly by solving $N$ 
single-particle Schr\"odinger-like equations, yielding the KS orbitals, 
from which the kinetic energy
may be constructed.   However, the KS DFT is not quite in keeping with 
the spirit of the Hohenberg-Kohn theorems,~\cite{HK} which provide a mathematical 
justification for the solution to the many-body problem solely in terms 
of the density of the system ({\it i.e.,} an orbital free
formulation).  Indeed, in its purest form, DFT has no need for the 
calculation of orbitals of any kind.  However, an orbital free DFT 
requires the specification of a noninteracting kinetic energy density 
functional which is not known exactly.
The purpose of this section is to provide an approximate, but accurate, kinetic energy 
density functional to be used in a DFT description of the ground state 
properties of a 2D harmonically trapped dipolar Fermi gas.

The first level of approximation for the explicit construction of the 
kinetic energy density functional is the local-density
approximation~\cite{DFT}.
In this approach, the exact kinetic energy density for a {\em 
homogeneous} system is determined, after which the same
expression is assumed to be true locally for the {\it inhomogeneous} 
system.  The LDA is generally valid for systems that are only weakly 
inhomogeneous, but may be remarkably accurate even for strongly 
inhomogeneous systems~\cite{brack_vanzyl,vanzyl5}.  In the case of a uniform
2D system, the noninteracting kinetic energy for a spin-polarized 2D Fermi 
gas is found to be
\beq\label{2dkehomo}
E^{\rm hom}_{K} = \sum_\bk n_\bk \frac{\hbar^2k^2}{2m}
= A \frac{\pi \hbar^2}{m} \bar{n}_{\rm 2D}^2~.
\eeq
Here $n_\bk = \theta(k_F-k)$ is the zero-temperature Fermi
occupation number, $k_F$ is the 2D Fermi wavevector given by $k_F^2 =
4\pi\bar n_{2D}$, $\bar{n}_{\rm 2D}$ is the uniform number 
density and $A$ is the area of the system.
Invoking the LDA, the 2D kinetic energy 
functional for an inhomogeneous system takes the form
\beq\label{2dkeLDA}
E_K[n] = \int d\br \frac{\pi\hbar^2}{m} n(\br)^2~.
\eeq
If to this we add the energy $E_P[n]$ of the particles
interacting with an external potential $v_{\rm ext}(\br)$, we
obtain the Thomas-Fermi (TF) energy functional
\beq
E_{TF}[n] = \int d\br \frac{\pi\hbar^2}{m} n(\br)^2 + \int d\br
\, v_{\rm ext}(\br) n(\br)~.
\label{E_TF}
\eeq
A variational minimization of this equation with respect to 
the density leads to the Euler-Lagrange equation
\beq\label{variation}
\frac{\delta E_{TF}[n]}{\delta n(\br)} - \mu_{TF} = 0~,
\eeq
where the Lagrange multiplier $\mu_{TF}$ (TF chemical
potential) serves to fix the total number of particles.
Using Eq.~(\ref{E_TF}), Eq.~(\ref{variation}) leads to the the TF spatial density, given explicitly as
\beq\label{TFden}
n_{TF}(\br) = \frac{m}{2\pi\hbar^2}\left (\mu_{TF} - v_{\rm
ext}(\br) \right )~.
\eeq
The density is seen to vanish on the surface defined by $v_{\rm
ext}(\br) = \mu_{TF}$ and is taken to be zero for all positions
where $v_{\rm ext}(\br) > \mu_{TF}$. This unphysical behaviour of
the density is of course a consequence of the local nature of
the TF energy functional, and is also present in other spatial
dimensions~\cite{brack_bhaduri}.

One may try to remove the unphysical behaviour of the TF density 
by improving upon the quality of the kinetic 
energy functional. One possibility is introducing 
gradient corrections which take into account
inhomogeneities of the density. Corrections of this kind can be
developed in several ways~\cite{kirzhnits57,HK,jones71,
hodges73,holas91,brack_bhaduri,PhD,koivisto,salasnich,
putaja,ghosh}. For example, one can consider a weakly
inhomogeneous system in which the density deviates only slightly
from some uniform value~\cite{HK}. However, 
if this method is used in 2D,
it is found that {\em all gradient corrections vanish}~\cite{
holas91,stern}.
The implication of this is that one cannot formally justify the 
inclusion of gradient corrections in 2D on the basis
of a systematic expansion about the homogenous limit.
Other methods, such as the semiclassical Wigner-Kirkwood (WK)
expansion~\cite{PhD,brack_bhaduri}, do yield 
gradient corrections in 2D which, however, make
no contribution to the total kinetic energy for physically
smooth densities~\cite{footnote1}. These observations of course 
do not mean that the LDA is exact in 2D
since the TF approximation certainly does not generate the exact
density of an inhomogeneous system. It is thus clear that 
nonlocal corrections to the
kinetic energy functional are necessary, and it is plausible that
they may still take the form of gradient corrections~\cite{ghosh}, 
similar to what are found in 1D and 3D inhomogeneous
systems~\cite{holas91}.  

To demonstrate the need for nonlocal corrections in
2D explicitly, it is useful to consider a 2D 
gas of noninteracting particles trapped within
the harmonic confining potential $v_{\rm ext}(r) = m\omega_0 
r^2/2$.  It was shown by Brack and van Zyl~\cite{brack_vanzyl} that
the exact spatial density is given by  
\beq\label{rhor_2D}
n_{\rm ex}(r) = \frac{1}{\pi} \sum_{n=0}^M (-1)^n (M-n+1) L_n(2r^2)
e^{-r^2}~,
\eeq
where $L_n(x)$ is a Laguerre polynomial~\cite{grad}, and all
lengths are expressed in units of
the harmonic oscillator length $a_{\rm ho} = \sqrt{\hbar/m\omega_0}$.
The integer, $M$, counts the number of filled oscillator shells, {\it i.e.,} the Fermi energy is given by $E_F=\hbar\omega_0 (M+1)$.  Integrating Eq.~(\ref{rhor_2D}) 
over all space leads to the total number of particles in the system as a function of the shell index
\beq\label{num2d}
N(M) = \frac{1}{2}(M+1)(M+2)~.
\eeq
The exact total energy of this system is found to be
\beq\label{KEexact}
E_{\rm ex} = \frac{\hbar\omega_0}{3} N \sqrt{1 + 8N}~.
\eeq
By virtue of the equipartition of the kinetic and potential
energies in a harmonic trap, the exact
kinetic energy is given by
\bea\label{KEexact_asymp}
E_{\rm ex}^K &=& \frac{\hbar\omega_0}{6} N \sqrt{1 +
8N}\nonumber \\ 
&=& \frac{1}{2} \hbar \omega_0 \left [
\frac{2 \sqrt{2}}{3} N^{3/2} + \frac{1}{12\sqrt{2}} 
N^{1/2} + O(N^{-1/2}) \right]~,
\eea
where in going to the last line in Eq.~(\ref{KEexact_asymp}), we have made use of the large-$N$ behaviour of the exact kinetic energy.

The remarkable result found in Ref.~\cite{brack_vanzyl} is that the 
LDA kinetic energy functional, Eq.~(\ref{2dkeLDA}), integrates 
to the {\it exact} kinetic energy when the exact density
is used as input. As mentioned above, this should not be misconstrued to mean that
Eq.~(\ref{2dkeLDA}) is in fact the exact kinetic energy 
functional for the harmonically trapped system, {\it i.e.} that 
no corrections beyond the LDA are required. 
The exact spatial density $n_{\rm ex}$ is emphatically not the 
density which variationally minimizes the TF energy functional.
This density is given by Eq.~(\ref{TFden}) and may be written as
\beq\label{2dTF}
n_{TF}(\br) = 
\frac{1}{4\pi a_{\rm ho}^4}\left(R_{TF}^2-r^2\right)~,
\eeq
where $R_{TF} = \sqrt{2\mu_{TF}/m\omega_0^2}$ 
is the TF radius, and $\mu_{TF} = \sqrt{2N}\hbar\omega_0$.
When Eq.~(\ref{2dTF}) is used in Eq.~(\ref{2dkeLDA}), the kinetic 
energy evaluates to
\beq\label{KELDA}
E_K[n_{TF}] = \frac{\sqrt{2}}{3} \hbar\omega_0 N^{3/2}~,
\eeq
which is the leading term in the large-$N$ expansion of $E_{\rm
ex}^K$.
Thus,
while Eq.~(\ref{2dkeLDA}) gives the quantum mechanical kinetic energy 
with the exact density as input, if the true variational
density, Eq.~(\ref{2dTF}), is used instead, the resulting kinetic 
energy is always {\em lower} than the exact kinetic energy.  
The upshot of all of this is that the TF energy functional, 
Eq.~(\ref{E_TF}), will always produce a kinetic energy that is lower 
than the true value. This implies that the kinetic energy part
of the TF functional has to be augmented by some correction 
in order to account for
the second term in Eq.~(\ref{KEexact_asymp})~\cite{footnote2}.  

We shall take the nonlocal correction to have the familiar 
von Weisz\"acker (vW) form~\cite{vW,brack_bhaduri}
\beq\label{vW}
E_{\rm vW}[n] =  \lambda_{\rm vW}(N) \frac{\hbar^2}{8 
m}\int d\br \frac{|\nabla n(\br)|^2}{n(\br)}~,
\eeq
where $\lambda_{\rm vW}(N)$ is a parameter which in general
can depend on the particle number, $N$. This functional has the
following desirable properties:
(i) it depends on the gradient of the spatial density
and thus vanishes in the limit of a uniform system, (ii) it is
positive definite, thus increasing the kinetic energy relative
to the TF approximation, and (iii) it scales in the same way as $E_K[n]$
so that equipartition of kinetic and potential energy is preserved. The total energy
functional for a noninteracting system in the TFvW approximation then reads (henceforth we 
suppress the $N$-dependence of $\lambda_{\rm vW}(N)$)
\beq\label{2dkevW}
E[n] = \int d\br \left[\frac{\pi\hbar^2}{m} n(\br)^2 + 
\lambda_{\rm vW}\frac{\hbar^2}{8m}\frac{|\nabla 
n(\br)|^2}{n(\br)}+ v_{\rm ext}(\br) n(\br)\right]~.
\eeq
The variational minimization of this energy functional is
conveniently achieved by introducing the so-called von
Weisz\"acker wave function $\psi(\br) \equiv \sqrt{n(\br)}$. The
Euler-Lagrange equation then takes the form of a nonlinear Schr\"odinger
equation,
\beq\label{SE_vW}
-\lambda_{\rm vW}\frac{\hbar^2}{2m} \nabla^2 \psi(\br) + v_{\rm 
eff}(\br) \psi(\br) = \mu \psi(\br)~,
\eeq
where the effective potential is given by
\beq\label{veff_vW}
v_{\rm eff}(\br) = \frac{2\pi\hbar^2}{m}\psi(\br)^2 + v_{\rm 
ext}(\br)~.
\eeq
Since $v_{\rm eff}(\br)$
itself depends on $\psi(\br)$, the solution of Eq.~(\ref{SE_vW})
must be determined self-consistently. The ground state solution
$\psi_0(\br)$ with the normalization
\beq\label{particles}
\int d\br |\psi_0(\br)|^2 = N~,
\eeq
determines the self-consistent ground state density $n_0(\br) = \psi_0(\br)^2$ 
and the chemical potential $\mu$.
We now establish that the vW correction can account for the
exact energy of the harmonically confined system with a
parameter $\lambda_{\rm vW}$ which is weakly $N$-dependent.

\begin{figure}[t]
\centering \scalebox{0.5}
{\includegraphics[34,120][500,490]{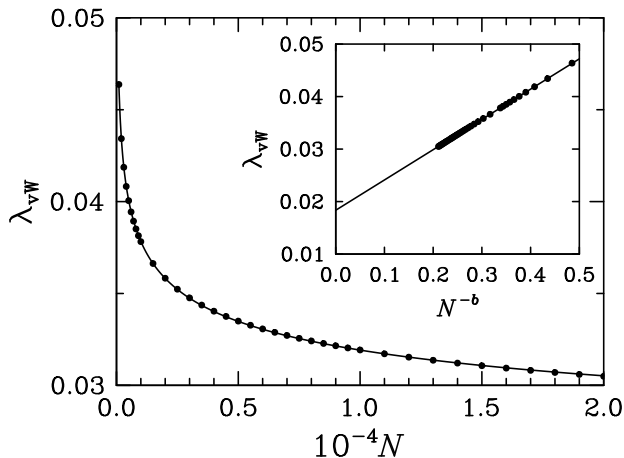}}
\caption{The von Weisz\"acker parameter $\lambda_{\rm vW}$ {\it
vs.} the number of particles $N$. The points are the
calculated values and the smooth curve is the fit
$\lambda_{\rm vW} = \lambda_{\rm vW}^\infty +a/N^b$ 
with $\lambda_{\rm vW}^\infty = 0.0184$, $a=0.0577$ and
$b=0.1572$. The inset illustrates the extrapolation $N\to
\infty$.
}
\label{lambda}
\end{figure}

For a given number of particles $N$ and a given value of
$\lambda_{\rm vW}$, we determine $n_0(\br)$ by solving the
closed set of equations, {\it viz.,} Eqs.~(\ref{SE_vW}) and
(\ref{veff_vW}), using the numerical method discussed in
Sec.~IV. This density is then used to evaluate
$E[n_0]$ and the result is compared to $E_{\rm ex}$. The parameter 
$\lambda_{\rm vW}$ is then adjusted and the procedure is
repeated until we achieve the equality
\beq\label{vWequal}
E[n_0] = E_{\rm ex}~.
\eeq 
The net result of this procedure leads to the values of 
$\lambda_{\rm vW}$ plotted as a function of $N$ in 
Fig.~\ref{lambda}. The $N$-dependence is indeed
weak and suggests that $\lambda_{\rm vW} \simeq 0.02$-0.04
for $N$ in the range $10^2$-$10^6$. These values of $\lambda_{\rm
vW}$ are considerably smaller than the value ($\sim 0.25$) found
to be appropriate in three-dimensions~\cite{chizmeshya88}. The inset to
Fig.~\ref{lambda} illustrates the extrapolation to $N\to \infty$
and demonstrates that $\lambda_{\rm vW}$ has a nonzero
limiting value.

To see in more detail how the vW energy accounts for the higher
order terms in Eq.~(\ref{KEexact_asymp}), it is convenient to
expand $E[n_0]$ in terms of the difference $\Delta n \equiv n_0
-n_{TF}$. We have
\bea\label{Ediff1}
E[n_0] &=& E_{TF}[n_0]+E_{\rm vW}[n_0] \nonumber \\
&=& E_{TF}[n_{TF}] + \frac{1}{2}m\omega_0^2 \int_{r \ge R_{TF}}
d\br (r^2-R_{TF}^2)\Delta n(\br) + \frac{\pi\hbar^2}{m} \int
d\br [\Delta n(\br)]^2 + E_{\rm vW}[n_0]~.
\eea
Equating (\ref{Ediff1}) to the large-$N$ expansion of
Eq.~(\ref{KEexact}) and using $E_{TF}[n_{TF}] =
(2\sqrt{2}/3)\hbar\omega_0 N^{3/2}$, we obtain
\beq
\frac{1}{12\sqrt{2}}\hbar \omega_0 N^{1/2}
\simeq E_{\rm vW}[n_0] +
\frac{1}{2}m\omega_0^2 \int_{r \ge R_{TF}} d\br
(r^2-R_{TF}^2)n_0(\br) + \frac{\pi\hbar^2}{m} \int d\br
[n_0(\br)-n_{TF}(\br)]^2~.
\label{E_vW_approx}
\eeq
Each of the integrals on the right hand side, including the
one defining $E_{\rm vW}[n_0]$, have integrands
that peak near $r \sim R_{TF}$. Thus the $N^{1/2}$
term in Eq.~(\ref{KEexact_asymp}), which is a correction to the
TF kinetic energy, can be interpreted as an edge correction. Since the
last two terms on the right hand side of Eq.~(\ref{E_vW_approx})
are small in comparison to $E_{\rm vW}[n_0]$, an approximate relation
determining $\lambda_{\rm vW}$ would be
\beq 
E_{\rm vW}[n_0] = \lambda_{\rm vW} \frac{\hbar^2}{8m} \int d\br
\frac{|\nabla n_0|^2}{n_0} \simeq \frac{1}{12\sqrt{2}}\hbar \omega_0
N^{1/2}~.
\label{E_vW_approx_2}
\eeq
In applying this relation we again note
that $n_0$ in Eq.~(\ref{E_vW_approx_2}) is
an implicit function of $\lambda_{\rm vW}$.  Thus, the procedure
described earlier is followed and $\lambda_{\rm vW}$ is adjusted
until $E_{\rm vW}[n_0]$ is equal to
the right-hand-side of Equation (\ref{E_vW_approx_2}).  
This yields values of $\lambda_{\rm vW}$ 
which are about 10\% larger than those obtained directly from
Equation (\ref{vWequal}).

\begin{figure}[t]
\centering \scalebox{0.5}
{\includegraphics[50,200][570,610]{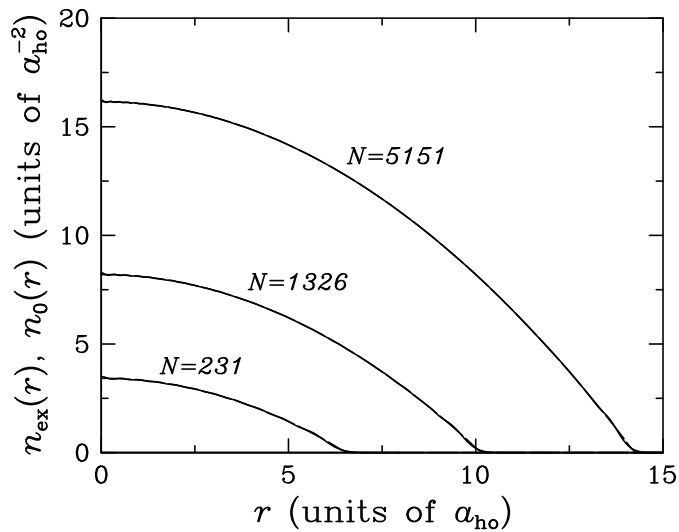}}
\caption{The exact ($n_{\rm ex}$, solid line) and
self-consistent
($n_0$, dashed line) densities as a function of the
radial distance $r$ for different numbers of particles $N$.  Even for
small particle numbers, the exact and TFvW densities are in very
good agreement.
}
\label{density}
\end{figure}

\begin{figure}[t]
\centering \scalebox{0.5}
{\includegraphics[50,200][570,610]{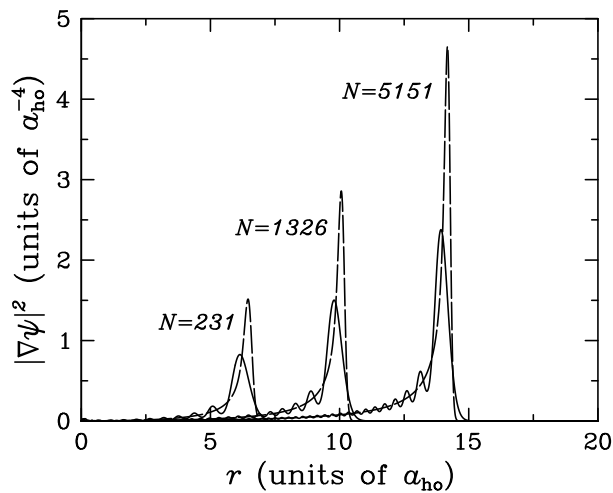}}
\caption{The quantity $|\nabla \psi|^2 = \frac{1}{4}|\nabla
n|^2/n$ contributing to the von Weisz\"acker kinetic energy
density in Eq.~(\ref{vW}), evaluated using the exact (solid 
line) and self-consistent (dashed line) spatial densities
as a function of the
radial distance $r$ for different numbers of particles $N$.
}
\label{vWKE}
\end{figure}

To the extent that $\lambda_{\rm vW}$ is weakly $N$-dependent, 
Eq.~(\ref{E_vW_approx_2}) indicates that the integral $\int d\br
|\nabla n_0|^2/n_0$ also scales roughly as $N^{1/2}$. This
$N$-dependence is also exhibited by the integral with $n_0$
replaced by $n_{\rm ex}$ which is an indication that these
densities are rather similar. In Fig.~\ref{density} we plot the 
self-consistent (solid line) and TFvW (dashed line)
densities for different particle numbers. Even for the relatively 
small value of $N=231$, these densities are in very good
agreement and in fact are very close to $n_{TF}$,
except at the edge of the cloud. The differences between $n_0$
and $n_{\rm ex}$ are more
clearly revealed by plotting $\frac{1}{4}|\nabla
n|^2/n=|\nabla \psi|^2$ in Fig.~\ref{vWKE}. 
The integrals of the curves shown in Fig.~\ref{vWKE} typically 
differ by about 15\%.
We note that the curves for the exact density (solid lines) exhibit
prominent oscillations which are associated with the orbital
shell structure. The fact that the TFvW curves (dashed lines) do not
exhibit this shell structure is entirely expected in view of
the semiclassical nature of the TFvW approximation~\cite{brack_bhaduri} . However,
we observe that the TFvW curves provide a smooth average of the shell
oscillations (a well known feature of semiclassical theories~\cite{brack_bhaduri})
up to the edge of the cloud. At the edge, the TFvW approximation 
overestimates the exact value. In principle, this discrepancy
can be reduced by including higher order gradient corrections 
to the kinetic energy functional, but for our purposes, this
refinement is not neccessary.

We summarize this section by noting that our analysis has shown that the corrections to the 2D 
LDA kinetic energy functional can indeed be represented in the
gradient vW form, in spite of the 
fact that gradient expansion methods fail to produce any such
terms in 2D. By comparing the results of the TFvW energy
functional to the exact results for a harmonically confined
noninteracting 2D gas, we are able to determine the vW
parameter $\lambda_{\rm vW}$ and show that it is weakly
$N$-dependent. We cannot claim that these values of
$\lambda_{\rm vW}$ are generally applicable for problems in 2D
but we would expect them to be appropriate for densities which are
similar to those of the harmonically confined system. In
particular, we expect the vW functional to be applicable for
a harmonically confined system in which interactions are also
included, as discussed in the following section.

\section{Dipolar Interactions: Hartree-Fock approximation}

Having developed the approximate kinetic energy functional in the previous section, we are now in a
position to construct the 
energy functional
which accounts for the dipolar interactions in a 2D
spin polarized Fermi gas. The approach we adopt is essentially
the one used for the analogous problem with Coulomb
interactions in 2D degenerate electronic systems. However, as we shall see, dipolar interactions lead
to a fundamentally different energy functional. At the level of
the Hartree-Fock (HF) approximation~\cite{DFT}, the
energy of the spin polarized Fermi gas
is $E_{HF} = E_K + E_{dd}$ where $E_K=\sum_\bk n_\bk
\hbar^2k^2/2m$ is the noninteracting kinetic energy and the
dipole interaction energy is
\beq
E_{dd} = \frac{1}{2}\sum_{\bk\bk'}n_\bk n_{\bk'}\left [ \langle
\bk \bk'|V_{dd}|\bk \bk'\rangle - \langle \bk \bk'|V_{dd}|\bk'
\bk \rangle \right ].
\label{E_dd}
\eeq
The first matrix element in Eq.~(\ref{E_dd})
is the direct term and the second is the exchange term. For the
singular dipolar interaction of interest ({\it i.e.,} 
with the spins polarized perpendicular to the plane),
\beq
V_{dd}(r) = \frac{\mu_0\mu^2}{4\pi r^3},
\label{dipole_int}
\eeq
each of these matrix elements are separately infinite.
However, the sum of the two terms is finite as a result of the
Pauli exclusion principle. This is seen most readily by writing
the interaction energy as
\beq
E_{dd} = \frac{A}{2}\int d^2r V_{dd}(\br) \bar n_{2D}^2 g_{HF}(\br),
\label{E_dd_2}
\eeq
where  the HF radial distribution 
function is defined as
\beq
\bar n_{2D}^2 g_{HF}(\br) = \bar n_{2D}^2 - \left|\frac{1}{A}\sum_\bk
e^{-i\bk\cdot\br}\right |^2.
\label{g_HF}
\eeq
Evaluating Eq.~(\ref{g_HF}), we find
\beq
g_{HF}(\br) = 1 - \left ( \frac{2J_1(k_F r)}{k_Fr} \right )^2,
\eeq
where $J_1(x)$ is the cylindrical Bessel function of order
one~\cite{grad}.
The function $g_{HF}(r)-1$ has a range of approximately
$k_F^{-1}$ and defines the `exchange hole' commonly used
in electron-gas theory.
For small $r$, $g_{HF}(r) \sim \frac{1}{4} (k_Fr)^2$, and as a
result, the integral in Eq.~(\ref{E_dd_2}) is well-behaved. We
find
\beq
E_{dd} =
\frac{1}{4}A\mu_0\mu^2\bar n_{2D}^2k_F \int_0^\infty dt \frac{1}{t^2}
\left [ 1 - \left ( \frac{2J_1(t)}{t^2} \right )^2\right ]
= A\frac{64}{45\sqrt{\pi}}\mu_0\mu^2\bar n_{2D}^{5/2}.
\label{E_dd_3}
\eeq
Equation (\ref{E_dd_3}) has also been derived in the paper by
Bruun and Taylor~\cite{bruun}, and
the final form is given by Fang and Englert~\cite{fang}. In particular, 
Eq.~(\ref{E_dd_3}) suggests that
one can define an interaction energy functional within
a LDA having the following form:
\beq
E_{dd}^{\rm LDA}[n] = \int d^2r \frac{2}{5}C_{dd} [n(\br)]^{5/2},
\label{E_dd_LDA}
\eeq
with $C_{dd} \equiv (32/9\sqrt{\pi})\mu_0\mu^2$. In contrast to
the Coulomb problem, this local functional presumably accounts
for the {\it total} interaction energy. To see if this is indeed
reasonable, one must investigate in more detail the effect of
density inhomogeneities.

Before doing so, we first
provide an alternative derivation of Eq.~(\ref{E_dd_3})
which assumes that $V_{dd}(r)$ has a well-defined Fourier
transform (FT). This is achieved by defining a {\it regularized} dipole
interaction $V_{dd}^{\rm reg}(r)$ which does not have the $r=0$
singularity of $V_{dd}(r)$. As we shall see, a regularized
interaction facilitates the corresponding analysis for an 
inhomogeneous system. Accepting for the moment that such
an interaction is available, Eq.~(\ref{E_dd}) becomes
\beq
E_{dd}^{\rm reg} = \frac{A}{2}\int \frac{d^2k}{(2\pi)^2}\int
\frac{d^2k'}{(2\pi)^2}\theta(k_F-k)\theta(k_F-k')\left [
\tilde V_{dd}^{\rm reg}(0)- \tilde V_{dd}^{\rm
reg}(\bk-\bk')\right ]~,
\label{E_dd_FT}
\eeq
where $\tilde V_{dd}^{\rm reg}(\bk)$ is the two-dimensional FT of
$V_{dd}^{\rm reg}(r)$. With the change of variable $\bk' =
\bk-\bq$, Eq.~(\ref{E_dd_FT}) can be written as
\beq
E_{dd}^{\rm reg} = \frac{A}{2}\int \frac{d^2q}{(2\pi)^4}
\left [ \tilde V_{dd}^{\rm reg}(0)- \tilde V_{dd}^{\rm
reg}(\bq)\right ]
\int d^2k\,\theta(k_F-k)\theta(k_F-|\bk+\bq|))
\label{E_dd_FT_2}~.
\eeq
The second integral is is just the area of overlap in momentum
space of two circles of radius $k_F$ whose centres are separated
by $q$. This area is given by
\beq
A(q,k_F) = 2k_F^2 \left [ \cos^{-1}\left(\frac{q}{2k_F}\right )
- \frac{q}{2k_F}\sqrt{1-\left(\frac{q}{2k_F}\right )^2}\right ]
\theta(2k_F-q)~.
\eeq
Substituting this result into Eq.~(\ref{E_dd_FT_2}), we find
\beq
E_{dd}^{\rm reg} = \frac{Ak_F^4}{2\pi^3}\int_0^1 dx \,x
\left [\cos^{-1}x - x\sqrt{1-x^2}\right ] \left [ \tilde V_{dd}^{\rm reg}(0)-
\tilde V_{dd}^{\rm reg}(2k_Fx)\right ]~.
\label{E_dd_FT_3}
\eeq
We now wish to show that this expression reduces to
Eq.~(\ref{E_dd_3}) using an appropriate limiting procedure.

To this end, we must now specify the regularized dipole interaction.
This is done by considering the interaction between two physical
{\it electric} dipoles~\cite{footnote3}, each of which has a charge distribution 
of the form
\beq
\rho(\br) =
\frac{2\bp\cdot\br}{\pi^{3/2}\sigma^5}e^{-r^2/\sigma^2}~.
\eeq
The electrostatic interaction between two dipoles $\bp_1$ and
$\bp_2$ separated by $\br$ is
\bea
U(\br)& = &\frac{1}{\epsilon_0}\int \frac{d^3k}{(2\pi)^3}
e^{i\bk\cdot\br}
\frac{(\bp_1\cdot\bk)(\bp_2\cdot\bk)e^{-k^2\sigma^2/2}}{k^2}
\nonumber \\
&=&-\frac{(\bp_1\cdot\nabla)(\bp_2\cdot\nabla)}{4\pi\epsilon_0}\left [
\frac{1}{r}{\rm Erf}\left(\frac{r}{\sqrt{2}\sigma}\right )
\right ]~,
\label{dipole_pot}
\eea
where ${\rm Erf}(x)$ is the error function.
For $r\gg \sigma$, this interaction reduces to that of two {\it
point} dipoles, varying as $r^{-3}$. However, for $r\ll \sigma$,
the interaction saturates at a constant value as a result of the
overlap of the dipole charge distributions.

Equation (\ref{dipole_pot}) can now be used to define a regularized
{\em magnetic} dipole interaction for the spin-polarized Fermi gas by
choosing $\bp_1=\bp_2 = p\hat{\bf z}$, putting $\br = (x,y,0)$
in Eq.~(\ref{dipole_pot}),
and replacing $p^2/\epsilon_0$ by
$\mu_0\mu^2$.  The regularized magnetic dipole interaction then reads
\bea
V_{dd}^{\rm reg}(x,y) &=& \mu_0\mu^2\int \frac{d^3k}{(2\pi)^3}
e^{i(k_xx+k_yy)}\frac{k_z^2}{k^2}e^{-k^2\sigma^2/2}\nonumber\\
&=& \frac{\mu_0\mu^2}{4\pi}\left [ \frac{1}{r^3}{\rm
Erf}\left(\frac{r}{\sqrt{2}\sigma}\right
)-\sqrt{\frac{2}{\pi}}\frac{1}{\sigma r^2}e^{-r^2/2\sigma^2} \right]~.
\label{V_dd_regularized}
\eea
Equation (\ref{V_dd_regularized}) approaches Eq.~(\ref{dipole_int}) 
for $r\gg \sigma$ and saturates at
$\mu_0\mu^2/3(2\pi)^{3/2}\sigma^3$ for $r\to 0$.
The two-dimensional FT of $V_{dd}^{\rm reg}(x,y)$ is
\bea
\tilde V_{dd}^{\rm reg}(q) &=& \mu_0\mu^2e^{-q^2\sigma^2/2}
\int \frac{dk_z}{2\pi}\frac{k_z^2}{k_z^2+q^2}
e^{-k_z^2\sigma^2/2}\nonumber \\
&=&\frac{\mu_0\mu^2}{\sqrt{2\pi}\sigma}\left
[e^{-q^2\sigma^2/2}-\sqrt{\frac{\pi}{2}}q\sigma {\rm
Erfc}\left(\frac{q\sigma}{\sqrt{2}}\right ) \right ]~.
\label{V_dd_regularized_FT}
\eea
For $\sigma k_F \ll 1$, we see that
\beq
\tilde V_{dd}^{\rm reg}(0)- \tilde V_{dd}^{\rm reg}(2k_Fx)\simeq
\mu_0\mu^2k_Fx~,
\eeq
to leading order in $\sigma k_F$. Inserting this result into
Eq.~(\ref{E_dd_FT_3}) we obtain
\beq
\lim_{\sigma \to 0} E_{dd}^{\rm reg} = A \frac{2\mu_0\mu^2 k_F^5}{45\pi^3}~,
\eeq
which is identical to Eq.~(\ref{E_dd_3}). We thus see that the
dipolar interaction energy can be obtained by taking the $\sigma\to 0$
limit in a calculation using
an appropriately defined regularized dipole interaction. Of course 
the definition of the regularized potential is not unique, but the
form we have chosen has particularly convenient properties. The
essential reason for being able to use this approach is that the
final result is insensitive to the
cutoff $\sigma$ when it becomes much smaller
than the range $k_F^{-1}$ of the exchange
hole. This calculation can be viewed as the momentum-space version
of the real-space approach leading to Eq.~(\ref{E_dd_2}).

We next proceed to a calculation of $E_{dd}$ for an arbitrary
{\em inhomogeneous} system making use of the regularized dipole
interaction defined above.
As we shall see, our real-space formulation leads
to a final result that is identical to that obtained by Fang
and Englert~\cite{fang} using a Wigner function representation.
However, our complementary derivation provides some additional 
insight into the
interaction energy functional of the dipolar Fermi gas.

The generalization of Eq.~(\ref{E_dd_2}) to an inhomogeneous
spin-polarized system is
\beq
E_{dd} = \frac{1}{2}\int d^2r \int d^2r'\left [
\rho(\br,\br)\rho(\br',\br') - \rho(\br,\br') \rho(\br',\br) 
\right ] V_{dd}(\br-\br')~,
\label{E_dd_general}
\eeq
where we have introduced the single particle density matrix
defined as
\beq
\rho(\br,\br') = \sum_{i,{\rm occ}} \phi^*_i(\br)\phi_i(\br')~.
\eeq
Here, the $\phi_i(\br)$ are a set of single-particle states that
correspond to a physical situation in which the
density $n(\br) = \rho(\br,\br)$ is localized in space.
The density matrix has the symmetry property $\rho(\br,\br') =
\rho(\br',\br)$.
The expression for $E_{dd}$ in Eq.~(\ref{E_dd_general}) is
well-defined even for the singular dipole interaction, however,
it is more difficult to exhibit the explicit cancellation between 
the direct and exchange terms for an inhomogeneous system. To
achieve this cancellation we make use of the regularized
interaction in Eq.~(\ref{V_dd_regularized}) which allows us to
evaluate the the direct and exchange terms separately.
The desired result is then obtained by taking the $\sigma \to 0$
limit at the end of the calculation. As we will show, the
singular parts of the direct and exchange terms  do in fact cancel
exactly.  

The direct term is calculated most
conveniently in momentum space. We have
\beq
E_{dd}^{(\rm d)} = \frac{1}{2}\int \frac{d^2q}{(2\pi)^2}
\tilde V_{dd}^{\rm reg}(q)|\tilde n(q)|^2~.
\label{E_d}
\eeq
To evaluate the exchange term, we introduce the centre-of-mass
variable $\bR = (\br+\br')/2$ and the relative variable $\bs
= \br-\br'$. The exchange term can then be written as
\beq
E_{dd}^{(\rm x)} = -\frac{1}{2}\int d^2s V_{dd}^{\rm reg}(s)\int d^2R
[\bar \rho(\bR,\bs)]^2~,
\label{Edd_x}
\eeq
where
\beq
\bar \rho(\bR,\bs) \equiv
\rho(\bR+\tfrac{1}{2}\bs,\bR-\tfrac{1}{2}\bs)~.
\eeq
The symmetry property of $\rho(\br,\br')$ implies that $\bar
\rho(\bR,-\bs) = \bar \rho(\bR,\bs)$.
We now define the function
\beq
f(\bs) = \int d^2R[\bar \rho(\bR,\bs)]^2~,
\label{f_s}
\eeq
which satisfies $f(-\bs) = f(\bs)$, and write
\bea
E_{dd}^{(\rm x)} &=& -\frac{1}{2}\int d^2s\, V_{dd}^{\rm reg}(s)f(\bs)
\nonumber \\
&=& -\frac{1}{2}\int d^2s\, V_{dd}^{\rm reg}(s)[f(\bs)-f(0)] -
\frac{1}{2}\int d^2s\, V_{dd}^{\rm reg}(s)f(0)~.
\label{E_x}
\eea
From Eq.~(\ref{f_s}) we have
\beq
f(0) =  \int d^2R [n(\bR)]^2 = \int \frac{d^2q}{(2\pi)^2}|\tilde
n(\bq)|^2~,
\eeq
where the last equality follows from Parseval's theorem, and we
have noted that $n(\bR) \equiv \bar\rho(\bR,0)=\rho(\bR,\bR)$.
Combining Eq.~(\ref{E_x}) and Eq.~(\ref{E_d}), we obtain
\beq
E_{dd} = \frac{1}{2}\int \frac{d^2q}{(2\pi)^2}
\left [ \tilde V_{dd}^{\rm reg}(q)-\tilde V_{dd}^{\rm reg}(0)
\right ]|\tilde n(\bq)|^2 -\frac{1}{2}\int d^2s\, V_{dd}^{\rm
reg}(s)[f(\bs)-f(0)]~.
\label{Edd_reg}
\eeq
It should be emphasized that this result is valid for any
potential $V_{dd}^{\rm reg}(r)$ which has a well-defined FT in
the $q\to 0$ limit. In case the potential does not have a 
well-defined $q\to 0$ FT, one must revert to the expressions in
Eqs.~(\ref{E_d}) and (\ref{Edd_x}).

Making use of Eq.~(\ref{V_dd_regularized_FT}) and taking
the $\sigma \to 0$ limit, we obtain
\bea
E_{dd} &=& -\frac{\mu_0\mu^2}{4}\int \frac{d^2q}{(2\pi)^2}
q\,|\tilde n(\bq)|^2 +\frac{\mu_0\mu^2}{8\pi}\int d^2s\,
\frac{1}{s^3} [f(0)-f(\bs)]\nonumber \\
&\equiv& E_{dd}^{(2)} + E_{dd}^{(1)}~.
\label{Edd_general}
\eea
It can be shown that Eq.~(\ref{Edd_general}) is identical to 
the result obtained by Fang and Englert~\cite{fang} and for this 
reason we
have adopted their notation for the two terms. The term
$E_{dd}^{(2)}$ appears explicitly in their paper but their
$E_{dd}^{(1)}$ is given in a different form, being
expressed in terms of the Wigner distribution
function. 

The $E_{dd}^{(1)}$ term involves the quantity
\bea
f(0)-f(\bs) &=& \int d^2R\left \{ [\bar \rho(\bR,0]^2 - [\bar
\rho(\bR,\bs]^2 \right \} \nonumber \\
&\equiv& \int d^2R [n(\bR)]^2 g(\bs,\bR)~.
\eea
Here we have defined the radial distribution function for the
inhomogeneous system as
\beq
g(\bs,\bR) = 1 - \frac{[\bar\rho(\bR,\bs)]^2}{[\bar
\rho(\bR,0)]^2}~.
\eeq
An approximation to $E_{dd}^{(1)}$ can be generated
by making a local
approximation for $g(\bs,\bR)$, namely,
\beq
g(\bs,\bR) \simeq g_{HF}(s;n(\bR))~,
\label{local_approx}
\eeq
where the radial distribution function of the uniform gas is evaluated
for a density equal to $n(\bR)$. With this replacement, we have
\beq
E_{dd}^{(1)} \simeq E_{dd}^{\rm LDA}~,
\eeq
as defined in Eq.~(\ref{E_dd_LDA}). The total interaction
includes the manifestly nonlocal $E_{dd}^{(2)}$ term.

We can check the validity of the LDA by evaluating
Eq.~(\ref{Edd_general}) for a model inhomogeneous
system. Specifically, we once again appeal to the
ideal 2D Fermi gas confined
by an isotropic harmonic trapping potential.
This model is especially useful since the
density matrix can be obtained in closed form for an arbitrary
number of filled shells. Scaling all lengths by the harmonic
oscillator length, $a_{\rm ho}$, the one-particle density
matrix for $M+1$ filled shells is given by~\cite{vanzyl5}
\beq\label{rhoRs_2D}
\bar\rho(\bR,\bs) = \frac{1}{\pi}\sum_{n=0}^M (-1)^n L_n(2 R^2) 
e^{-R^2} L_{M-n}^1(s^2/2) e^{-s^2/4}~,
\eeq
where $L_n^\alpha(x)$ is the associated Laguerre
polynomial~\cite{grad}.
We observe that for this particular model system,
$\bar\rho(\bR,\bs)$ depends only on the magnitudes of the
vectors $\bR$ and $\bs$, which is also true of all quantities
derived from it.
Putting $s = 0$ in Eq.~(\ref{rhoRs_2D}) yields the density
$n(R)$ given by Eq.~(\ref{rhor_2D}). (For convenience we drop 
the `ex' subscript in the following.) 
The FT of $n(R)$ is readily found to be given by
\beq\label{nq_2D}
n(q) = L_M^2(q^2/2) e^{-q^2/4}~.
\eeq
We may also provide an explicit expression for the function $f(\bs)$, namely,
\bea\label{fns}
f(s) &\equiv&  \int d^2R |\bar\rho(\bR,\bs)|^2\nonumber\\
&=& \frac{1}{2 \pi} \sum_{n=0}^M \left[ L_{n}^1(s^2/2\right)]^2 e^{-s^2/2}~.
\eea
Since all of the functions required for the evaluation of 
Eq.~(\ref{Edd_general}) depend only on
the magnitude of the coordinates, the angular integrations can 
be performed immediately, leading to
\bea\label{Edd_2D_1}
E_{dd} &=& -\frac{\mu_0\mu^2}{8\pi a_{\rm ho}^3}\int_0^\infty dq~
q^2\,|\tilde n(q)|^2 +\frac{\mu_0\mu^2}{4 a_{\rm ho}^3}
\int_0^\infty \frac{ds}{s^2} [f(0)-f(s)]\nonumber \\
&=& -\frac{\mu_0\mu^2}{8\pi a_{\rm ho}^3}\int_0^\infty dq~
q^2\,|\tilde n(q)|^2 -\frac{\mu_0\mu^2}{4 a_{\rm ho}^3}
\int_0^\infty \frac{ds}{s} \frac{df(s)}{ds}~.
\eea
Using Eq.~(\ref{fns}) along with Eq.~(\ref{nq_2D}), we obtain
\bea\label{Edd_2D_2}
E_{dd}^{(1)} &=&
\frac{\mu_0\mu^2}{8\pi a_{\rm ho}^3} \sum_{n=0}^M
\left[\int_0^\infty ds~ 2 
L^2_{n-1}(s^2/2)L_{n}^1(s^2/2)e^{-s^2/2} +\int_0^\infty ds~[L_{n}^1(s^2/2)]^2 
e^{-s^2/2}\right]~.\label{Edd(1)}\\
E_{dd}^{(2)} &=&
-\frac{\mu_0 \mu^2}{8\pi a_{\rm ho}^3}\int_0^\infty dq~ q^2 [L_M^2 (q^2/2)]^2 
e^{-q^2/2}.
\label{Edd(2)}
\eea
The function $L_{-1}^2(x)$, arising from the $n=0$ term in the summation of Eq.~(\ref{Edd(1)}), should be
interpreted as zero.
All of the integrals in Eqs.~(\ref{Edd(1)}) and (\ref{Edd(2)})
can be evaluated 
analytically, using the following general result~\cite{vanzyl5}
\bea\label{Imn}
I_{mn}(\alpha,\beta,\gamma) &=& \int_0^{\infty} dx~ x^{\alpha} e^{-x} L^{\beta}_m(x) L^{\gamma}_n(x)\nonumber \\
&=& \frac{\Gamma(1+\alpha)\Gamma(n+\gamma+1)\Gamma(\beta-\alpha+m)}{\Gamma(m+1)\Gamma(n+1)\Gamma(1+\gamma)\Gamma(\beta-\alpha)}~
_3F_2\left(1+\alpha-\beta,-n,1+\alpha;1+\gamma,1+\alpha-\beta-m;1\right)~,
\eea 
where $_3F_2(a,b,c;d,e;z)$ is the generalized hypergeometric
function~\cite{grad}.
A direct application of Eq.~(\ref{Imn}) gives
\begin{eqnarray}\label{Eddfinal}
&&\hskip -.5truein E_{dd}^{(1)} = \frac{\mu_0 \mu^2}{4 \pi a_{\rm
ho}^3}\frac{1}{\sqrt{2}} 
\sum_{n=0}^M
\frac{(n+1)\Gamma(n+3/2)}{\Gamma(n+1)}\left\{
\tfrac{4}{3} n~_3F_2 \left(-\tfrac{3}{2},\tfrac{1}{2},-n;
2,-\tfrac{1}{2}-n;1\right)
+~_3F_2 \left( -\tfrac{1}{2},\tfrac{1}{2},-n;
2,-\tfrac{1}{2}-n;1\right)\right\}, \\
&&\hskip -.5truein E_{dd}^{(2)} = 
-\frac{\mu_0 \mu^2}{4 \pi a_{\rm ho}^3}\frac{1}{2\sqrt{2}} \frac{\Gamma
\left(M+\frac{3}{2}\right) (M+1)(M+2)}{\Gamma(M+1)}~_3F_2
\left(\tfrac{1}{2},\tfrac{3}{2},-M;2,\tfrac{1}{2}-M;1\right)~.
\end{eqnarray}
While these final expressions are not particularly illuminating, 
they do provide summations which are easily dealt with numerically.   
On the other hand, the integrands in
Eqs.~(\ref{Edd(1)}) and (\ref{Edd(2)}) become highly oscillatory
when the index $n$ is large and the evaluation of the integrals
can be problematic 
without the use of specialized numerical integration techniques.
\begin{table}[ht] 
\centering      
\begin{tabular}{c c c c c}  
\hline\hline                        
$N$ & $E^{(1)}_{dd}$ & $E^{\rm LDA}_{dd}$ & $E_{dd}^{(2)}$& $\Delta E/E \%$ \\ [0.5ex] 
\hline                    
55    &  54.4003   &   54.5724   &   -6.8321   &   15\\
105   &  168.6535 &    168.9366  &   -15.3065   &   10\\
 231 &    670.199   &     670.718   &    -40.9666  &    7\\ 
 1326 &  14266.3& 14268.3 & -363.681 & 3\\ 
 5151 & 153345.6 & 153349.4 & -1983.06 & 1 \\[1ex]       
\hline     
\end{tabular} 
\caption{Comparison of the dipolar interaction energies with the LDA.  
The last column corresponds to the relative percentage error between 
the exact total energy, $E^{\rm exact}_{dd} = 
E^{(1)}_{dd}+E^{(2)}_{dd}$, and 
$E^{\rm LDA}_{dd }$. The unit
of energy is $\mu_0\mu^2/a_{\rm ho}^3$.} 
\label{TableI}  
\end{table} 

\begin{figure}[t]
\centering \scalebox{0.5}
{\includegraphics[50,200][570,610]{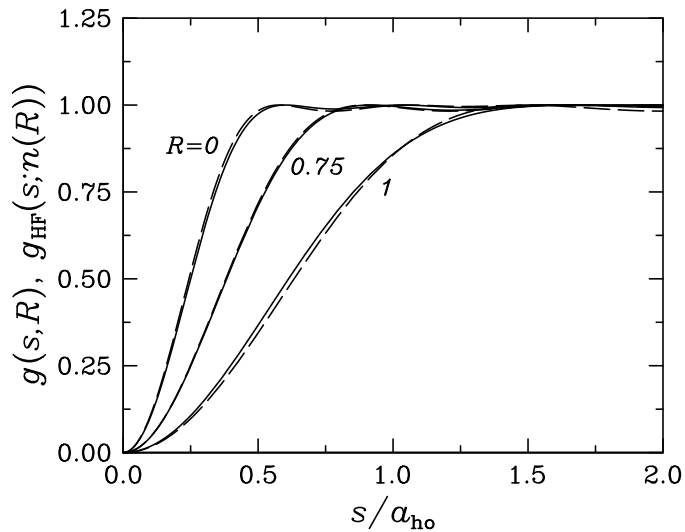}}
\caption{Comparison of the exact radial distribution function (solid)
with the local approximation (dashed). The values of $R$
indicated are in units of the TF radius $R_{TF}$. The number of
particles is $N=231$ ($M=20$).
}
\label{hole}
\end{figure}

In Table \ref{TableI} we give values of $E_{dd}^{(1)}$, 
$E_{dd}^{(2)}$ and $E_{dd}^{\rm LDA}$ for a range of particle 
numbers, $N$. We see that
$E_{dd}^{(1)} \simeq E_{dd}^{\rm LDA}$ to a very good
approximation even for relatively small numbers of particles.
To quantify this further we show in Fig.~\ref{hole} a comparison of 
$g_{HF}(s;n(R))$ and $g(s,R)$ for $M=20$ ($N=231$) for 
$R$ ranging from the 
centre of the cloud to its edge. We see that the local 
approximation in Eq.~(\ref{local_approx}) is
very good; the local Fermi wavevector $k_F(R)$ captures very
nicely the extent of the exchange hole in the exact radial
distribution function. These observations explain why 
$E_{dd}^{(1)}$ is so close to $E_{dd}^{\rm LDA}$. Although
these results have been obtained for the specific model of a
harmonically confined gas, we would expect similar
behaviour whenever the model density varies on a length
scale which is large compared to the extent of the local
exchange hole. 

Table \ref{TableI} also shows that the nonlocal contribution 
diminishes rapidly with respect to the local contribution with
increasing $N$.  
Indeed, it is straightforward to show that the $N\gg 1$ behaviour
of the two contributions are $E^{(1)}_{dd} \sim N^{7/4}$ and 
$|E^{(2)}_{dd}| \sim N^{5/4}$.  Therefore, we find
that  $|E_{dd}^{(2)} / E_{dd}^{(1)}| \sim 1/\sqrt{N}$ in the 
large-$N$ limit. 
As a result, the {\em total} interaction energy is very well 
approximated by $E_{dd}^{\rm LDA}$ for large $N$, as illustrated in Table \ref{TableI}.
This local energy functional for the total interaction
energy can thus be trusted in applications, such as those
typically encountered in traps, where the density of
the system is a smooth and slowly varying function of position.

We wish to emphasize that the locality of the interaction energy
functional is a property of the dipolar interaction and is not
generally valid. To illustrate this we can compare these results
with those obtained for an interparticle Coulomb interaction,
$e^2/4\pi\epsilon_0 r$. There
is no need to regularize the potential in this case and the
direct and exchange terms can be evaluated directly from
Equation (\ref{E_dd_general}). For consistency we again consider a
spin-polarized situation. The direct contribution is
\beq
E^{\rm (d)}_{ee} = \frac{e^2}{8\pi \epsilon_0}\int 
\frac{d^2q}{(2\pi)^2} \frac{2\pi}{q} |n(\bq)|^2~,
\eeq
where $2\pi/q$ is the two-dimensional FT of $r^{-1}$.
The exchange term reads
\beq
E_{ee}^{\rm (x)} = -\frac{e^2}{8\pi \epsilon_0}\int 
\frac{d^2s}{s}f(\bs)~.
\eeq
Using Eqs.~(\ref{nq_2D}) and (\ref{fns}) for 
the isotropic two-dimensional harmonic trap, we find
\beq
E_{ee}^{\rm (d)} = \frac{e^2}{4\pi\epsilon_0a_{\rm ho}}
\frac{1}{3 \sqrt{2}} 
\frac{\Gamma\left(M+\frac{5}{2}\right) \Gamma(M+3)}
{\Gamma(M+1)^2}~_3F_2
\left(-\tfrac{3}{2},\tfrac{1}{2},-M;3,-\tfrac{3}{2}-M;1\right)~,
\eeq
and
\beq\label{exch_ex}
E_{ee}^{\rm (x)} = -\frac{e^2}{4\pi\epsilon_0a_{\rm ho}} 
\frac{1}{\sqrt{2}} \sum_{n=0}^M \frac
{\Gamma\left(n+\frac{3}{2}\right) (n+1)} {\Gamma(n+1)}~_3F_2
\left(-\tfrac{1}{2},\tfrac{1}{2},-n;2,-\tfrac{1}{2}-n;1\right)~.
\eeq
The sum of these energies can be thought of as an approximation
to the interaction energy of a 2D parabolic quantum dot.
The 2D (spin-polarized) Dirac exchange functional in the LDA is given by
\beq\label{dirac_ex}
E^{\rm (x),LDA}_{ee}[n] = -\frac{e^2}{4\pi\epsilon_0a_{\rm ho}}
\frac{8}{3\sqrt{\pi}}\int d^2 r [n(r)]^{3/2}. 
\eeq

We compare the various energies in Table~\ref{TableII}. We see
that the LDA is again a very good approximation to the exchange
energy. The direct Coulomb energy is of course inherently
nonlocal and is seen to give the dominant interaction energy
contribution for large $N$, in contrast to the situation for the
dipolar interaction where the local contribution dominates.  
In fact, it can readily be shown that $|E^{\rm (x)}_{ee}/E^{\rm (d)}_{ee}|\sim 1/\sqrt{N}$ for $N\gg 1$.
\begin{table}[ht]
\centering 
\begin{tabular}{c c c c c} 
\hline\hline 
$N$ & $E^{\rm (x)}_{ee}$ & $E^{\rm (x),LDA}_{ee}$ & $E_{ee}^{\rm
(d)}$&
$\Delta E/E \%$ \\ [0.5ex] 
\hline 
55  &  -85.8544 &  -85.4033  & 683.615   &  0.5\\
105  & -192.347 &  -191.746 &  2119.36 &  0.3\\
231 &  -514.802 &  -513.958  & 8421.97  &  0.2\\
1326 & -4570.15 & -4568.39  & 179276.2 & 0.04\\
5151 & -24919.8 &  -24911.7 & 1.92699 $\times 10^6$  & 0.03\\
[1ex] 
\hline  
\end{tabular}
\caption{Comparison of the Coulomb interaction energy with the
exchange energy calculated exactly and in the LDA.
The last column gives the relative percentage error
between the exact exchange energy, Eq.~(\ref{exch_ex}), and the
exchange energy obtained from the Dirac functional, Eq.~(\ref{dirac_ex}). The energies
are in units of $e^2/4\pi \epsilon_0 a_{\rm ho}$.} 
\label{TableII} 
\end{table}

To summarize, we have shown that even for modest values of the 
particle number, $N$, the total interaction energy for the 
harmonically trapped, 2D dipolar Fermi gas can be approximately 
represented by a {\em local}
energy functional.  We must emphasize again that the locality of the density functional for the dipolar gas 
is a result of the short range $\sim 1/r^3$ behaviour of the dipole-dipole interaction, in contrast to {\it e.g.,}
the Coulomb interaction, in which case the {\em total} interaction energy cannot be reduced to a purely local form.

We now have all of the necessary components to construct the TFvW energy functional for a 2D, harmonically trapped
interacting dipolar Fermi gas, {\it viz.,}
\beq
\label{ETFvW}
E[n] = \int d\br \left[\frac{1}{2}C_K n(\br)^2 + 
\lambda_{\rm vW}\frac{\hbar^2}{8m}\frac{|\nabla 
n(\br)|^2}{n(\br)}+ \frac{2}{5}C_{dd} [n(\br)]^{5/2}+v_{\rm
ext}(\br) n(\br)\right],
\eeq
where $C_K = 2\pi\hbar^2/m$.
Once again, introducing the vW wavefunction, and performing the
variational minimization of Eq.~(\ref{ETFvW}) with respect to the 
density, gives 
\beq\label{SE_TFvW}
-\lambda_{\rm vW}\frac{\hbar^2}{2m} \nabla^2 \psi(\br) + v_{\rm 
eff}(\br) \psi(\br) = \mu \psi(\br)~,
\eeq
where now, $v_{\rm eff}(\br)$ also contains an interaction term, {\it viz.,}
\beq\label{veff_TFvW}
v_{\rm eff}(\br) = C_K\psi(\br)^2 + C_{dd} \psi(\br)^3 + v_{\rm 
ext}(\br)~.
\eeq
Along with the normalization condition, Eq.~(\ref{particles}), Eq.~(\ref{SE_TFvW}) provides a complete description for the ground state spatial density
of the system.

\section{Self-consistent equilibrium solutions}

The numerical self-consistent solution of Eq.~(\ref{SE_TFvW})
was achieved by means of imaginary time propagation (ITP)
\cite{trotter}. In this method, the time-dependent Schr\"odinger 
equation 
\begin{equation}
\label{itp}
\dfrac{\partial \psi(\br,\tau)}{\partial\tau} = - \dfrac{H(\tau)}{\hbar} 
\psi(\br,\tau)~,
\end{equation}
is evolved in imaginary time, $t \to -i \tau$, starting from some
arbitrary initial state $\psi(\br,0)$. 
The Hamiltonian governing the evolution is
\beq\label{Ham}
H(\tau) = -\lambda_\text{vW} \dfrac{\hbar^2\nabla^2}{2m} 
+ v_\text{eff}(\br,\tau) \equiv T +V(\tau)~,
\eeq
where the time dependence arises from the dependence of
$v_\text{eff}$ on the evolving density $n(\br,\tau) =
\psi(\br,\tau)^2$. The evolution is carried out in a step-wise
fashion according to
\begin{equation}\label{psi_imag}
\psi(\br,\tau + \Delta\tau) = e^{-H(\tau)\Delta\tau/\hbar}
\psi(\br,\tau)~.
\end{equation}
The repeated application of the evolution operator
yields a wave function $\psi(\br,\tau)$ which eventually converges 
to the self-consistent ground state $\psi_0(\br)$ as $\tau
\to \infty$.

The evolution in Eq.~(\ref{psi_imag}) is achieved by using
the split-operator approximation~\cite{trotter,strang}
\begin{equation}
e^{-H(\tau)\Delta\tau/\hbar} \simeq e^{-V(\tau)\Delta\tau/2\hbar}
e^{-T\Delta\tau/\hbar} e^{-V(\tau)\Delta\tau/2\hbar}~,
\end{equation}
together with fast Fourier transforms (FFT)~\cite{numerical} to convert between
coordinate and momentum spaces. If a Cartesian grid is used 
in the $x$ and $y$ directions for our two-dimensional geometry,
2D FFTs are required. However, a more efficient algorithm is
available if the system possesses circular symmetry.
The kinetic energy operator in this case takes the form
\begin{equation}
T = -\lambda_{\rm vW}\frac{\hbar^2}{2m} \frac{d^2}{dr^2} +
\lambda_{\rm vW}\frac{\hbar^2}{2m} \frac{1}{r}\frac{d}{dr}\equiv
T_1+T_2~.
\end{equation}
With this decomposition, the evolution operator becomes
\begin{equation}
\label{tesplit}
e^{-H(\tau)\Delta\tau/\hbar} \simeq e^{-V(\tau)
\Delta\tau/2\hbar} e^{- T_2 \Delta\tau/2\hbar} e^{-T_1 
\Delta\tau/\hbar} e^{- T_2 \Delta\tau/2\hbar} e^{- 
V(\tau) \Delta\tau/2\hbar}~,
\end{equation}
where the $T_1$ step is again treated by means of FFTs, but now
with respect to the one-dimensional radial variable $r$ in the
range $[-R,R]$. The kinetic step $T_2$, on the other hand, is
treated in coordinate space using a Crank-Nicholson method~\cite{numerical} which
requires the solution of a tridiagonal system, {\it viz.},
\begin{equation}\label{phi}
-\phi(r_{i+1}) + \alpha_i \phi(r_i) + 
\phi(r_{i-1}) = \bar\phi(r_{i+1}) + \alpha_i 
\bar\phi(r_i) - \bar\phi(r_{i-1})~,
\end{equation}
where $\alpha_i = \dfrac{4 r_i}{\lambda_\text{vW}} \dfrac{\Delta 
r}{\Delta\tau}$, and $\bar\phi(r_i)$ is the wave function
prior to the application of the $T_2$ evolution operators.
The solution to Eq.~(\ref{phi}) is uniquely determined by 
Dirichlet boundary conditions at the ends of the $r_i$ mesh,
where the wave function is required to vanish. 
At the end of each time step we update $H$ with the new $n(r,\tau)$ 
which is properly normalized to $N$.
The convergence criterion for achieving self-consistency 
is taken to be $\sum_i 
|\psi(r_i,\tau_{n}) - \psi(r_i,\tau_{n-1})| < \varepsilon$, where 
typically $\varepsilon \lesssim 10^{-6}$. Once self-consistency
has been achieved,
the chemical potential is given by $\mu =
\langle \psi_0| H | \psi_0\rangle$ and the ground state energy
is obtained from Equation (\ref{ETFvW}).
This numerical procedure can also be adapted with minor 
modifications to spherically symmetric 3D systems.

\begin{figure}[t]
\centering \scalebox{0.5}
{\includegraphics[50,200][570,610]{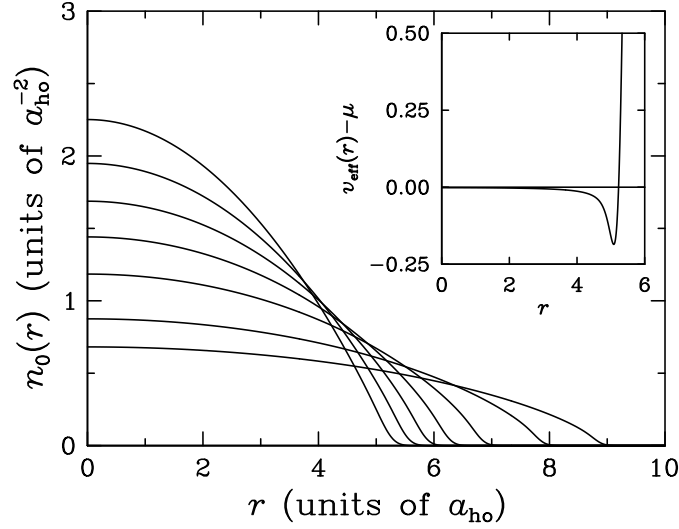}}
\caption{Density distributions for $N=100$. The curves with
increasing radial extent correspond to $C_{dd}/C_K = 0$, 0.2,
0.5, 1.0, 2.0, 5.0, 10.0.  The inset shows the effective potential in
units of $\hbar \omega_0/2$, as a function of $r$ in units of $a_{\rm ho}$,
in the noninteracting limit, $C_{dd}/C_K=0$.
}
\label{den_TFDW_N100}
\end{figure}

\begin{figure}[t]
\centering \scalebox{0.5}
{\includegraphics[50,200][570,610]{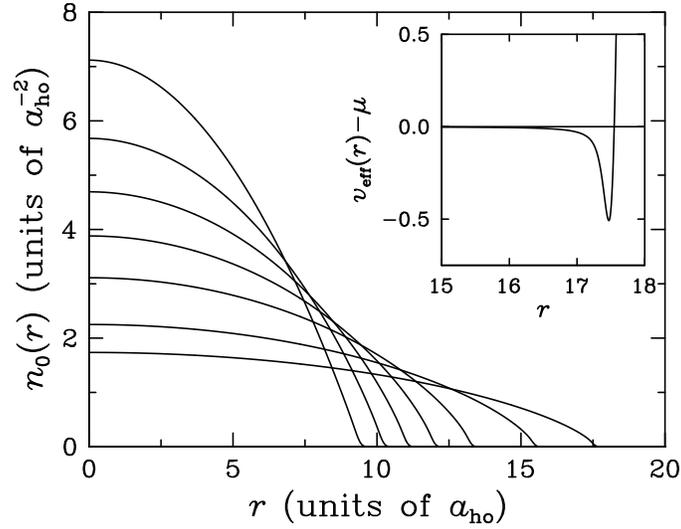}}
\caption{Density distributions for $N=1000$. The curves with
increasing radial extent correspond to $C_{dd}/C_K = 0$, 0.2,
0.5, 1.0, 2.0, 5.0, 10.0.  The inset shows the effective potential in
units of $\hbar \omega_0/2$, as a function of $r$ in units of $a_{\rm ho}$,
in the case of strong interactions, $C_{dd}/C_K=10.0$.
}
\label{den_TFDW_N1000}
\end{figure}

In Fig.~\ref{den_TFDW_N100} we present the TFvW self-consistent 
ground state density profile for $N=100$ for various strengths 
of the dipole-dipole interaction. With increasing interaction
strength, the radius of the atomic cloud increases and the
central density decreases, which is expected given the repulsive nature
of the dipole-dipole interaction for the spin-polarized case. As found for the $C_{dd} =0$
case considered earlier, the vW gradient term has the effect of
giving a density that decays smoothly in the classically
forbidden region defined by $v_{\rm eff}(r) > \mu$.
We recall that if the vW term is absent, the spatial density has 
an unphysical sharp cut-off at a radius $R$ defined by $v_{\rm
eff}(R) = \mu'$, where $\mu'$ is the chemical
potential in this approximation. For $C_{dd}/C_K \gg 1$, $R
\propto N^{3/10}$.
In Fig.~\ref{den_TFDW_N1000}, we show the spatial density for 
$N=1000$. It is apparent that the effect of the vW
term becomes less significant as the number of particles is increased.
This trend increases with increasing $N$ and the density
approaches the distribution found in the local density
approximation. In the inset to both Figs.~\ref{den_TFDW_N100}
and \ref{den_TFDW_N1000} we have also included representative
plots of the effective potential for the case $C_{dd}/C_K=0$ in
Fig.~\ref{den_TFDW_N100} and $C_{dd}/C_K=10$ in
Fig.~\ref{den_TFDW_N1000}.  The potential is essentially flat up
to the edge of the cloud where it falls below the chemical 
potential and then rises quadratically. The main point
to be taken from these curves is that the shape of the effective 
potential is not very sensitive to the introduction of interactions. 
Finally, we note that our equilibrium density
profiles are very similar to those found in Ref.~\cite{fang}
without the vW correction. The latter densities would be virtually
indistinguishable from those plotted in Figs. 5 and 6
except in the classically forbidden region where the
vW gradient correction leads to densities which decay smoothly
to zero.

\section{Conclusions}
We have presented a mathematically simple, and computationally 
efficient DFT formulation of the
equilibrium properties of a 2D trapped dipolar Fermi gas based on the 
Thomas-Fermi von Weizs\"acker approximation.  
One of the key elements of this work is the development of a
kinetic energy functional appropriate for an inhomogeneous
2D system.  Specifically, we have shown that the addition of a 
vW-like gradient correction to the TF kinetic energy functional 
is needed in order to
accurately determine the ground state energy, as well as to
provide a physically 
reasonable density distribution at the edge of the cloud.
While conventional gradient expansions in 2D fail to yield gradient 
corrections to the TF kinetic energy functional, our 
detailed analysis has provided a compelling 
argument for the inclusion of a 
vW-like gradient correction.
We have also provided an alternative derivation of the interaction
energy functional first obtained by Fang and 
Englert~\cite{fang}. We find in particular that the
exchange hole is a useful concept in formulating the local
density approximation, and furthermore explains
the underlying reasons behind the local nature 
of the total interaction functional.  

We have also presented a highly efficient self-consistent
numerical scheme for determining 
the equilibrium spatial density distributions within
the TFvW formalism. Our calculations
illustrate how the strength of the (repulsive) 
dipole-dipole interaction affects the size of the cloud and its
spatial distribution. Although the vW gradient
correction does not modify substantially the density in the
interior of the cloud, it does yield a density which decays
smoothly into the classically forbidden region 
at the edge. This feature is particularly
important in performing calculations of the collective mode 
frequencies using TFvW hydrodynamics~\cite{zaremba_tso,vanzyl4}, and
will be addressed in a future publication.
In addition, it will be of interest to investigate the affect of the anisotropy of the dipole-dipole
interaction in a confined 2D system by considering the 
situation in which the spins are canted at some angle with
respect to the 2D plane. Finally, although we have focused on
the 2D geometry, we wish to emphasize that the extension of the
TFvW theory to 3D is straightforward.

\acknowledgements
This work was supported by the Natural Sciences and Engineering Research Council of Canada (NSERC).

\end{document}